\documentclass[preprint,pra,showpacs,nofootinbib]{revtex4}
\def\be{\begin{equation}}
\def\ee{\end{equation}}
\def\ben{\begin{eqnarray}}
\def\een{\end{eqnarray}}

\begin{document}

\begin{flushright}
\vspace*{1cm}
\end{flushright}
\title{\bf Brownian motion of a charged test particle \\ near a reflecting boundary at finite temperature }

\author{  Hongwei Yu, Jun Chen  and  Puxun Wu }

\affiliation {Department of Physics and Institute of  Physics,\\
Hunan Normal University, Changsha, Hunan 410081, China}

\begin{abstract}

\baselineskip=18pt
 We discuss the random motion of charged test particles driven by quantum
electromagnetic fluctuations at finite temperature in both the
unbounded flat space and flat spacetime with a reflecting boundary
and calculate the mean squared fluctuations in the velocity and
position of the test particle.  We show that typically the random
motion driven by the quantum fluctuations is one order of
magnitude less significant than that driven by thermal noise in
the unbounded flat space. However, in the flat space with a
reflecting plane boundary, the random motion of quantum origin can
become much more significant than that of thermal origin at very
low temperature.

\end{abstract}

\maketitle \baselineskip=14pt


\section{Introduction }

 Quantum fluctuations, especially quantum vacuum
fluctuations, have been subjected to extensive studies, since the
emergence of quantum theory which has profoundly changed our
conception of empty space or vacuum. Two well-known examples of
experimentally verified effects resulting from changes of vacuum
fluctuations are the Lamb shift and the Casimir effect
\cite{Casimir,MR99,SKL99}. A fundamental feature to be expected of
any field which is quantized is the quantum fluctuations.
Therefore, test particles under the influence of these quantum
field fluctuations will no longer move on the classical
trajectories, but undergo random motion around a mean path.  It
will be very desirable and quite interesting to bring to light the
basic features of this kind of random motion driven by quantum, as
opposed to classical or thermal-like fluctuations.

In investigating the random motion  of test particles driven by
quantum field fluctuations, a natural first step is to examine the
case of vacuum, since, quantum-theoretically, quantum fields
fluctuate even in vacuum. However, because of the divergences that
arise in quantum field theory in unbounded Minkowski spacetime
when vacuum is concerned, it appears that the most tractable cases
of random motion of test particles in vacuum would be those in
which changes of vacuum fluctuations occur due to the  presence of
boundaries or non-trivial topology in a local flat space-time. The
simplest example of this is the random motion of a charged test
particle caused by {\it changes} in the electromagnetic vacuum
fluctuations near a perfectly reflecting plane boundary, which has
recently been investigated \cite{YF04}\footnote{Another example of
this quantum random motion is the random motion of photons due to
modified quantum fluctuations of the quantized gravitational
field~\cite{F95,YF99,YF00,YW}, which induces quantum lightcone
fluctuations.}. There, the effects have been calculated of the
modified electromagnetic vacuum fluctuations due to the presence
of the boundary upon the motion of a charged test particle . In
particular, it has been shown \cite{YF04} that the mean squared
fluctuations in  velocity and position of the test particle normal
to the plane can be associated with an effective temperature of
\be
   T_{eff}={\alpha\over \pi}\;{1\over k_B mz^2}=
  1.7\times 10^{-6}\left({{1 \mu m}\over z}\right)^2\;K
= 1.7\times 10^2\left({1\mathring{A}\over z}\right)^2\;K\; ,
\label{eq:effT}
 \ee
where $k_B$ is Boltzmann's constant and $z$ is the distance from
the boundary.  This might be experimentally accessible in the
future. These results have also been generalized to the case of
two parallel reflecting plates \cite{YC}.

As further step along the line, naturally, one would be interested
in a physically more interesting case, i.e.,  the random motion of
test particles caused by quantum field fluctuations at non-zero
temperature (as opposed to zero temperature vacuum fluctuations)
in the unbounded flat spacetime and flat spacetimes with
boundaries.
These are just what we want to address in the present paper. We
would like to study the random motion of a charged test particle
subject to ever-existing quantum electromagnetic fluctuations at
finite temperature,  i.e., the random motion driven by quantum
fluctuations of a thermal bath of photons. It will be demonstrated
that, for the random motion driven by quantum electromagnetic
field fluctuations at finite temperature, no dissipation is needed
for the velocity dispersion of the test particle to be bounded at
later times, in contrast to that driven by thermal noise.
Moreover, it will be shown that, in the unbounded flat spacetime,
generally the random motion driven by quantum fluctuations is one
order of magnitude less significant than that driven by thermal
noise. However, it could be strengthened if the quantum field
fluctuations are to be modified by the presence of a reflecting
plane boundary and even become orders of magnitude more
significant than that of thermal origin, when the system
temperature is low.


\section{ Brownian motion of the test particle in Minkowski space at finite temperature}


First, let us now consider the motion of a charged test particle
subject to quantum electromagnetic field fluctuations at finite
temperature $T$ in the Minkowski (unbounded flat) space. We will
use Lorentz-Heaviside units with $c=\hbar =1$ in our discussions.
 In the limit of small velocities, the motion of a charged
particle is described by a non-relativistic equation of motion
(Langevin equation) with a fluctuating electric force
\begin{equation}
\frac{d {\mathbf v}}{d t}= \frac{e}{m}\,
{\mathbf{E}}({\mathbf{x}},t)\,;
 \label{eq:Langevin}
\end{equation}
 assuming that the
particle is initially at rest and has a charge to mass ratio of $
e/m$. The velocity of the charged particle at time $t$ can be
calculated as follows
 \begin{equation}
 {\mathbf{v}}={e\over m}\int_0^t\;{\mathbf{E}}({\mathbf{x}}\;,
t)\;dt =\bigg({4\pi\alpha\over
m^2}\bigg)^{1/2}\int_0^t\;{\mathbf{E}}({\mathbf{x}}\;, t)\;dt \;,
\end{equation}
where $\alpha$ is the fine-structure constant.
 The mean squared fluctuations in speed
in the $i$-direction can be written as (no sum on $i$)
\begin{eqnarray}
\langle{\Delta v_i^2}\rangle &=&{4\pi\alpha\over
m^2}\;\int_0^t\;\int_0^t\; \langle{E}_i({\mathbf
x},t_1)\;{E}_i({\mathbf x},t_2)\rangle_\beta\,dt_1dt_2
 \; ,   \label{eq:lang2}
\end{eqnarray}
where $ \langle{E}_i({\mathbf x},t_1)\;{E}_i({\mathbf
x},t_2)\rangle_\beta $ is understood to be the renormalized
electric field two-point function at finite temperature $T={1\over
k_B \beta}$ and we have used the fact that $\langle E_i
\rangle_{\beta}=0$. We adopt the well-established renormalization
procedure in quantum field theory in which physical quantities are
calculated and supposedly experimentally measured against vacuum.
Therefore, the renormalized electric field two-point function is
obtained by subtracting the vacuum contribution. We have, for
simplicity, assumed that the distance does not change
significantly on the time scale of interest in a time $t$, so that
it can be treated approximately as a constant. If there is a
classical, nonfluctuating field in addition to the fluctuating
quantum field, then Eq.~(\ref{eq:lang2}) describes the velocity
fluctuations around the mean trajectory caused by the classical
field. Note that when the initial velocity does not vanishes, one
 has to also consider the influence of fluctuating magnetic fields
 on the velocity dispersion of the test particles. However, it has
 been shown that this influence is, in general, of
the higher order than that caused by fluctuating electric fields
and is thus negligible \cite{TY}.

 Let us note that the two point function for the photon
field at finite temperature, $ D_\beta^{\mu \nu} (x,x') = \langle
0| A^\mu (x)\, A^\nu (x') |0 \rangle_\beta$,  can be written as an
infinite imaginary-time image sum of the corresponding
zero-temperature two-point function, ${D^{\mu \nu}_{0}} (x-x')$,
i.e.,
\begin{equation}
\label{eq:TwoPoint}
 D_\beta^{\mu \nu} (x,x')  =\sum^{\infty}_{n=-\infty}\;{D^{\mu \nu}_{0}} (\mathbf{x-x'},t-t'+in\beta)
 \;,
\end{equation}
where argument $x$ stands for a four-vector, i.e.,  $(\;
\mathbf{x},t\; )$.
 In the Feynman gauge, we have
\begin{equation}
\label{eq:TwoPoint0}
D_{0}^{\mu\nu}(x-x')=\frac{\eta^{\mu\nu}}{4\pi^{2}(\Delta
t^{2}-\Delta \mathbf{x}^{2})}\;.
\end{equation}
By taking the four dimensional curl in $x$ and in $x'$, we can
obtain the electric field two-point function from that of the
photon field as follows
 \begin{eqnarray}
 \label{eq:curl}
 \langle E_{i}(x)E_{j}(x')\rangle=\langle
F_{0i}(x)F_{0j}(x')\rangle =\partial_{0}\partial_{0} ^{'} \langle
A_{i}(x)A_{j}(x')\rangle+\partial_{i}\partial_{j} ^{'} \langle
A_{0}(x)A_{0}(x')\rangle\;.
 \end{eqnarray}
The components of the renormalized electric field two-point
function at finite temperature, $\langle{\mathbf E}({\mathbf
x},t_1)\;{\mathbf E}({\mathbf x},t_2)\rangle_\beta$, can be
obtained by taking curl of  Eq.~(\ref{eq:TwoPoint}) according to
Eq.~(\ref{eq:curl}) and dropping the vacuum term ($n=0$ term in
the sum). The result is
\begin{eqnarray}
&&\langle
E_{x}({\mathbf{x}},t')E_{x}({\mathbf{x}},t'')\rangle_\beta
=\langle
E_{y}({\mathbf{x}},t')E_{y}({\mathbf{x}},t'')\rangle_\beta=\langle
E_{z}({\mathbf{x}},t')E_{z}({\mathbf{x}},t'')\rangle_\beta
\nonumber\\&&\quad =\frac{1}{\pi^{2}}\sum^{\infty}_{n=-\infty}{'}
{1\over (\Delta t + in\beta)^4}={\pi^2\over 3\beta^4}\bigg(2+
\cosh {2\pi\Delta t\over \beta}\bigg) \text {csch}
^4\biggl({\pi\Delta t\over \beta}\biggr)-{1\over \pi^2\Delta
t^4}\;. \label{eq:Ex}
\end{eqnarray}
Here a prime means that the $n=0$ term is omitted in the
summation. It is interesting to note that the first term in the
above result is the usual finite temperature correlation function
that satisfies the Kubo-Martin-Schwinger relation while the last
is the vacuum term (zero temperature contribution).  Therefore,
the renormalized correlation function does not obey the KMS
relation. Mathematically one can obtain a regularized correlation
function that satisfies the KMS relation by subtracting both the
$n=0$ and $n=1$ terms. However, the problem is that one does see
any physical motivation in removing the $n=1$ mode in contrast to
in deducting the $n=0$ one which amounts to taking away the vacuum
contribution. Let us also note here that the two-point
electromagnetic field correlation functions in black body
radiation have been examined in the literature, see for example,
Ref.~\cite{EK,BM69}.

Substituting the above results into Eq.~(\ref{eq:lang2}) and
carrying out the integration, we find that the velocity
dispersions  are given by
  \ben \label{eq:vx}
 \langle \Delta v_x^2\rangle&=&\langle \Delta v_y^2\rangle=\langle \Delta v_z^2\rangle={e^2\over
m^2}\;\int_0^t\;\int_0^t\;\langle{E}_x({\mathbf
x},t')\;{E}_x({\mathbf
 x},t'')\rangle_\beta\; dt'\;dt''\nonumber\\
 &=& {e^2 \text{csch}^2\bigl({\pi t\over
\beta}\bigr)\over
18\pi^2m^2\beta^2t^2}\biggl[5\pi^2t^2+3\beta^2+(\pi^2t^2-3\beta^2)\cosh
{2\pi t\over \beta} \biggr]\;.
 \een
 In the low temperature
limit, i.e., when $\beta \gg t$, we have
 \be
 \langle \Delta v^2\rangle=  \langle \Delta v_x^2\rangle +
 \langle \Delta v_y^2\rangle + \langle \Delta v_z^2\rangle={e^2\pi^2\over 15
m^2\beta^2}\bigg({t\over \beta}\bigg)^2- {2e^2\pi^4\over 189
m^2\beta^2}\bigg({t\over \beta}\bigg)^4 \;.
  \ee
This result shows that the velocity dispersion decreases very
quickly as inverse powers of $\beta ^4$ and it approaches zero
when $\beta\rightarrow \infty$ as expected.
 While in the high temperature limit, i.e., when $t\gg \beta$,
 \be
 \label{eq:vm}
 \langle \Delta v^2\rangle ={e^2\over 3 m^2\beta^2}-
{e^2\over \pi^2 m^2t^2} \;.
 \ee
To get a concrete idea of how large $t$ should be in order that
the condition $t\gg \beta$ is fulfilled, let us assume that the
temperature $T$ is about $\sim 10^2$ Kevin, which can well be
considered as high since we are discussing a quantum effect, then
the condition becomes $ t\gg 5.7\times 10 ^{-14} sec.$. This is
rather small. It is interesting to note that, for the random
motion driven by quantum fluctuations at finite temperature here,
no dissipation is needed for $\langle \Delta v_i^2\rangle$ to be
bounded at late times in contrast to the random motion due to
thermal noise.

 The mean squared position fluctuations  can
be calculated as follows
 \ben
 \langle\Delta x^2\rangle&=&\langle\Delta y^2\rangle=\langle\Delta z^2\rangle=
\int_0^t\;dt_1\;\int_0^{t_1}\;dt'\;\int_0^t\;dt_2
\;\int_0^{t_2}\;dt''\; \langle{E}_x({\mathbf
x},t')\;{E}_x({\mathbf
 x},t'')\rangle_\beta\nonumber\\
&=& {e^2\over 18\pi^2m^2\beta^2}\Bigg(\pi^2t^2-6\pi
t\beta\coth{\pi t\over
\beta}+6\beta^2\bigg[1+\ln\bigg({{\beta\over \pi t}}\sinh{\pi
t\over \beta}\bigg)\bigg]\Bigg)\;.
 \een
The limiting forms for both low and high temperature
approximations are respectively
 \be
 \langle\Delta x^2\rangle={e^2\pi^2t^4\over 180
m^2\beta^4}-{\pi^4t^6\over 1701m^2\beta^6 }\;, \quad\quad \beta\gg
t\;,
 \ee
and
  \be
  \label{eq:xm}
  \langle\Delta x^2\rangle={e^2t^2\over 18
m^2\beta^2}-{e^2\over 3\pi^2m^2 }\ln {\pi t\over \beta}+ {e^2\over
3\pi^2m^2 }\;, \quad\quad t\gg\beta\;.
 \ee
Eq.~(\ref{eq:xm}) reveals that $\sqrt{\langle\Delta x^2\rangle}$
grows linearly with time, and thus in principle can increase
indefinitely with time.  However, recall that we have assumed that
the particle do not move very far on the time scale of interest in
a time $t$. Therefore, it is quite compelling for us to figure out
under what conditions Eq.~(\ref{eq:xm}) is compatible with our
initial approximation, which disregards the displacement of the
particle. For this purpose, let us note that a natural time scale
of interest here is set by $\beta$, the inverse of the temperature
of the system. Hence, we expect our results to be a good
approximation as long as $\langle\Delta x^2\rangle\ll \beta^2$.
This equivalent to requiring that
 \be
 t \ll {3\over \sqrt{2\alpha\pi}}(m\beta )\beta\;.
 \ee
Note that $m\beta $ is the ratio of the temperature corresponding
to the mass of the particle to that of the system, which is
typically very large. Take an electron for example, the
temperature corresponding to the electron mass is $\sim 5.93\times
10^9$ K. Therefore, our results can be valid as long as the system
temperature is not any close to this value. This is expected to be
fulfilled by any experiment at the Earth. Finally, let us note
that this kind of random motion  driven by quantum fluctuations is
superimposed on that driven by thermal noise. Let the root mean
squared fluctuations in velocity due to the random motion driven
by quantum fluctuations be denoted by$ \Delta v_{qm} =
\sqrt{\langle\Delta v^2\rangle}$ and that by thermal noise at the
same temperature by $ \Delta v_{th}$, then it is easy to show that
 \be
 {\Delta v_{qm}\over \Delta v_{th}
 }={2\over 3}(\pi\alpha)^{1/2}\approx 0.1=10^{-1}\;.
 \ee
 This indicates that typically the random motion driven by
 quantum fluctuations is one order of magnitude less  significant than
 that driven by thermal noise.


\section{ Brownian motion of the test particle near a reflecting boundary at finite temperature}


Now a question arises naturally as to what happens if we modify
the quantum field fluctuations by adding a boundary in space, a
perfectly reflecting plane, for example. In particular, we are
interested in whether the random motion driven by quantum field
fluctuations at finite temperature will be strengthened or
weakened by the modification. Suppose such a reflecting plate be
located at the $z=0$ plane and the test particle be initially at a
distance $z$ from the plate, then the electric field two-point
function at finite temperature, $\langle{\mathbf E}({\mathbf
x},t_1)\;{\mathbf E}({\mathbf x},t_2)\rangle_\beta$,  can be found
by the method of double images with one involving an image source
displaced in the $z$-direction and the other involving an infinite
sum of temperature images displaced in imaginary time. At a point
a distance $z$ from the plane, the results are
 \ben
&&\langle
E_{x}({\mathbf{x}},t')E_{x}({\mathbf{x}},t'')\rangle_\beta
=\langle
E_{y}({\mathbf{x}},t')E_{y}({\mathbf{x}},t'')\rangle_\beta\nonumber\\
&&\quad =\sum^{\infty}_{n=-\infty}{'} {1\over \pi^2(\Delta t +
in\beta)^4)}-\sum^{\infty}_{n=-\infty} {(\Delta t
+in\beta)^2+4z^2\over \pi^2[(\Delta t +
in\beta)^2-4z^2)]^3}\nonumber\\
&&\quad \equiv F_{\beta m}(\Delta t,z )+ F^x_{\beta b}(\Delta t,
z)\;, \een
 and
 \ben
&&\langle
E_{z}({\mathbf{x}},t')E_{z}({\mathbf{x}},t'')\rangle_\beta\nonumber\\
&&\quad =\sum^{\infty}_{n=-\infty}{'} {1\over \pi^2(\Delta t +
in\beta)^4)}-\sum^{\infty}_{n=-\infty} {1\over \pi^2[(\Delta t +
in\beta)^2-4z^2)]^2}\nonumber\\
&&\quad \equiv F_{\beta m}(\Delta t,z )+ F^z_{\beta b}(\Delta t,
z)\;, \een
 where we have defined
 \ben
 F_{\beta m}(\Delta t,z )&=&{\pi^2\over 3\beta^4}(2+ \cosh {2\pi\Delta t\over \beta})
{\text {csch} } ^4\biggl({\pi\Delta t\over \beta}\biggr)-{1\over
\pi^2\Delta t^4}\; \een
 and
 \ben
 F^x_{\beta b}(\Delta t, z) &=&- {1\over 64 \pi\beta z^3}\biggl(\coth
{\pi(\Delta t-2z)\over \beta}-\coth {\pi(\Delta t+2z)\over
\beta}\biggr)\nonumber\\
&& +{1\over 32\beta^2z^2} \biggl(\text{csch} ^2{\pi(\Delta
t-2z)\over \beta}+\text{csch} ^2{\pi(\Delta t+2z)\over
\beta}\biggr)\nonumber\\
&& - {\pi\over 8\beta^3 z}\bigg(\coth {\pi(\Delta t-2z)\over
\beta}\text{csch}^2{\pi(\Delta t-2z)\over \beta}\nonumber\\
&&\quad\quad\quad\quad-\coth {\pi(\Delta t+2z)\over
\beta}\text{csch}^2{\pi(\Delta t+2z)\over
\beta}\, \bigg)\;,\nonumber\\
 \een
 \ben
 F^z_{\beta b}(\Delta t, z) &=&- {1\over 32 \pi\beta z^3}\biggl(\coth
{\pi(\Delta t-2z)\over \beta}-\coth {\pi(\Delta t+2z)\over
\beta}\biggr)\nonumber\\
&& +{1\over 16\beta^2z^2} \biggl(\text{csch} ^2{\pi(\Delta
t-2z)\over \beta}+\text{csch} ^2{\pi(\Delta t+2z)\over
\beta}\biggr)
 \een
Clearly, $F_{\beta m}$ is the electric field two-point function at
finite temperature in Minkowski space while $F^x_{\beta b}$ and
$F^z_{\beta b}$ are the correction induced by the presence of the
boundary. With the electric field two-point function given, the
velocity dispersion in the $x$-direction can be calculated out to
be
 \ben
 \langle \Delta v_x^2\rangle&=&{e^2\over m^2}\Bigg\{ \bigg( {1\over9\beta}-{1\over 3\pi^2t^2}+
 {\text {csch}^2{\pi t \over \beta}\over 3\beta^2}\bigg) -{1\over 16\pi^2z^2}\ln \bigg( {\sinh{\pi (t+2z) \over \beta}
 \sinh{\pi (2z-t) \over \beta}\over \sinh^2{2\pi z \over \beta}}
 \bigg)\nonumber\\
 &&\quad\quad-{1\over 4\pi\beta z}\coth{2\pi z\over \beta}\;\text{csch}{\pi(t -2z)\over
 \beta}\;
 \text{csch}{\pi(t+2z)\over \beta}\; \sinh^2{\pi t\over \beta}\nonumber\\
 &&\quad\quad+{\beta\over
 128\pi^3z^3}\bigg(g_\beta(t,z)-g_\beta(t,-z)\bigg)\Bigg\}\;.
 \een
Here we have introduced a new function $g_\beta(t,z)$, which is
defined by
 \be
g_\beta(t,z)=\text{PolyLog}[\;2, e^{2\pi(t-2z)\over
\beta}\;]+2\text{PolyLog}[\;2, e^{4\pi z\over \beta}\;]+
\text{PolyLog}[\;2, e^{-2\pi(t+2z)\over \beta}\;]\;,
 \ee
where the polylogarithm functions, PolyLog[n, z, ] are given by
 \be
\text{PolyLog}[\;n, z\;]= \sum_{k=1}^{\infty}\;{z^k\over
k^n}\equiv PL[\;n,z\;]\;.
 \ee
The velocity dispersion in the $z$-direction is given by
 \ben
 \langle \Delta v_z^2\rangle&=&{e^2\over m^2}\Bigg\{ \bigg( {1\over9\beta}-{1\over 3\pi^2t^2}+
 {\text {csch}^2{\pi t \over \beta}\over 3\beta^2}\bigg) -{1\over 8\pi^2z^2}\ln \bigg( {\sinh{\pi (t+2z) \over \beta}
 \sinh{\pi (2z-t) \over \beta}\over \sinh^2{2\pi z \over \beta}}
 \bigg)\nonumber\\
 &&\quad\quad+{\beta\over
 64\pi^3z^3}\bigg(g_\beta(t,z)-g_\beta(t,-z)\bigg)\Bigg\}\;.
 \een
The mean squared fluctuations in both the transverse and
longitudinal directions are evaluated to be
 \ben
\langle\Delta x^2\rangle&=&{e^2\over m^2}\Bigg({t^3\over 8\pi\beta
z^2}- {t^2\over 8\pi\beta z}\coth {2\pi z\over \beta}-{t^2\over
32\pi^2 z^2}\ln \bigg[4\sinh^2 {2\pi t\over \beta}\bigg]-{t\over
2\pi\beta}\nonumber\\
&&\; +{t\over 8\pi^2 z}\ln \bigg[\text{csch}{\pi(t-2z)\over
\beta}\sinh{\pi(t+2z)\over \beta}\bigg]\;\Bigg) +{e^2\over
m^2}\Bigg( {\beta(t^2-8z^2)\over 128\pi^3z^3}\text{PL}[2,
e^{-2\pi(t+2z)\over
\beta}] \nonumber\\
&&\;+ {\beta(t^2-8z^2)\over 128\pi^3z^3}\text{PL}[2, e^{4\pi
z\over \beta}]+{\beta t(t+4z)\over 128\pi^3z^3}\text{PL}[2,
e^{2\pi(t+2z)\over \beta}]+{\beta^2\over 64\pi^4z^2}\text{PL}[3,
e^{4\pi z\over \beta}]\nonumber\\
&&\;+{\beta^2(t-2z)\over 128\pi^4z^3}\text{PL}[3,
e^{2\pi(t-2z)\over \beta}]+{\beta^3\over
256\pi^5z^3}\bigg[\text{PL}[4, e^{-4\pi z\over
\beta}]+\text{PL}[4, e^{2\pi(t+2z)\over \beta}]\bigg]\nonumber\\
&&\;+ (z\rightarrow -z)\;\Bigg)\;,
 \een
and
 \ben
\langle\Delta z^2\rangle&=&{e^2\over m^2}\Bigg({t^3\over 8\pi\beta
z^2}-{t^2\over 32\pi^2 z^2}\ln \bigg[4\sinh^2 {2\pi t\over
\beta}\bigg]\;\Bigg) +{e^2\over m^2}\Bigg( {\beta t^2\over
64\pi^3z^3}\text{PL}[2, e^{-2\pi(t+2z)\over
\beta}] \nonumber\\
&&\;+ {\beta t^2\over 64\pi^3z^3}\text{PL}[2, e^{4\pi z\over
\beta}]+{\beta t(t+4z)\over 64\pi^3z^3}\text{PL}[2,
e^{2\pi(t+2z)\over \beta}]+{\beta^2\over 32\pi^4z^2}\text{PL}[3,
e^{4\pi z\over \beta}]\nonumber\\
&&\;+{\beta^2(t-2z)\over 64\pi^4z^3}\text{PL}[3,
e^{2\pi(t-2z)\over \beta}]+{\beta^3\over
128\pi^5z^3}\bigg[\text{PL}[4, e^{-4\pi z\over
\beta}]+\text{PL}[4, e^{2\pi(t+2z)\over \beta}]\bigg]\nonumber\\
&&\;+ (z\rightarrow -z)\;\Bigg)\;.
 \een
Here $(z\rightarrow -z)$ stands for all the terms in the big
brackets but with  the sign of $z$ flipped. In the high
temperature limit $t\gg z\gg \beta$, the velocity and position
dispersions of the test particle in the directions parallel to the
plate are approximately given by,
  \be
   \langle \Delta v_x^2\rangle=\langle \Delta v_y^2\rangle\approx {e^2\over 9
   m^2\beta^2}-{e^2\over 8\pi m^2\beta z}
+{e^2\beta \over128 \pi m^2 z^3}
   -{e^2\over
   3\pi^2m^2t^2}\;,
   \ee
and
 \be
\langle \Delta x^2\rangle=\langle \Delta y^2\rangle\approx
{e^2\over m^2}\left({t^2\over 18 \beta^2}-{t^2\over 16 \pi
z\beta}+{t\over 2\pi \beta}-{1\over 3\pi^2}\ln{\pi t\over
\beta}+{1\over 3\pi^2}\right)\;,
 \ee
while for the direction normal to the plate, we have
 \be
   \langle \Delta v_z^2\rangle\approx {e^2\over 9
   m^2\beta^2}+{e^2\over 4\pi m^2\beta z}
+{e^2\beta \over 64 \pi m^2 z^3}
   -{e^2\over
   3\pi^2m^2t^2}\;,
   \ee
\be \langle \Delta z^2\rangle\approx {e^2\over m^2}\left({t^2\over
18 \beta^2}+{t^2\over 8 \pi z\beta}+{2t\over \pi \beta}-{1\over
3\pi^2}\ln{\pi t\over \beta}+{1\over 3\pi^2}\right)\;.
 \ee
 Let us that note that, for $z=1\mu m$,  the condition $z\gg \beta$
leads to the requirement that the temperature of the system be
much larger only than $10^{-3}$ K. Hence, depending on the value
of $z$, a very low temperature in experiment may be considered as
high temperature for the random motion discussed here. A
comparison of the above results with Eq.~(\ref{eq:vm}) and
Eq.~(\ref{eq:xm} ) reveals that the random motion driven by
quantum field fluctuations at finite temperature is reinforced in
 the normal direction and weakened in the parallel directions by the
presence of a reflecting plate, which modifies the quantum field
fluctuations.  It is easy to see that even with this enhancement
the random motion in the normal direction driven by quantum
fluctuations is still much less significant than that driven by
thermal noise. This is expected since when the temperature is
high, the random motion should be dominated by thermal noise.

When the temperature of the system is very low, i.e., when
$\beta\gg t$ and $\beta\gg z$, in the $x$-direction, the
dispersions of the test particle are approximated as follows
 \be
 \langle \Delta v_x^2\rangle\approx {e^2\over \pi^2 m^2}\;\left[
{t\over 64z^3}\ln\biggl( {2z+t\over
 2z-t}\biggr)^2-{ t^2\over 8z^2(t^2-4z^2)}\right]+{32e^2\pi^4\over 945m^2}{t^2z^2\over
 \beta^6}\;,
 \ee
 \ben
 \langle\Delta x^2\rangle&\approx& \frac{e^2}{\pi^2m^2}\, \left[-{t^2\over 24z^2}+{t^3\over192z^3}\,
\ln\left(\frac{t+2z}{t-2z}\right)^2 -{1\over
6}\ln\left(\frac{t^2-4z^2}{4z^2}\right) \right]\nonumber\\
&& +{8\pi^4e^2\over 945 m^2}{t^4z^2\over \beta^6}\;,
 \een
and in the $z$-direction as follows
 \ben
 \langle \Delta v_z^2\rangle&\approx &{e^2\over \pi^2 m^2}\; {t\over 32z^3}\ln\biggl( {2z+t\over
 2z-t}\biggr)^2+{64e^2\pi^4\over 945m^2}{t^2z^2\over \beta^6}\;,
 \een
 \ben
 \langle \Delta z^2\rangle&\approx& \frac{e^2}{\pi^2 m^2}\, \left[\frac{t^2}{24 z^2}
+ \frac{t^3}{96 z^3}\, \ln\left(\frac{t+2z}{t-2z}\right)^2 +
\frac{1}{6}\,
\ln\left(\frac{t^2-4z^2}{4z^2}\right)\right]\nonumber\\
&& + {\pi^2t^4\over 90\beta^4}-{2\pi^4(5t^6+18t^4z^2)\over
8505\beta^6}\; ,
 \label{eq:z2}
 \een
 The $\beta$ independent terms in all the above expressions result
 from the Brownian motion driven just by the quantum vacuum fluctuations, while $\beta$ dependent terms
 represent the temperature corrections. When $\beta\rightarrow \infty$, the above results reduces to those given in
 Ref.~\cite{YF04} for the Brownian motion in vacuum. Clearly, in the low
 temperature limit, the Brownian motion is dominated by the
 quantum vacuum fluctuations and the temperature corrections are
 higher order and thus negligible. It is worth noting that, depending on the value of the initial distance of
 the test particle from the plate,  the
 temperature $T$  may have to be extremely low in order for the
 low temperature condition $\beta\gg z$ to be obeyed. For example,
 for $z=1\mu m$, the temperature $T$ must be lower than $10^{-3}$
 K. Therefore, in reality, we are more likely to face the high
 temperature limit, i.e., when $t\gg \beta$ and $z \gg \beta$ are
 satisfied.

However, if the system temperature is so low such that $\beta\gg
t\gg z$ holds, then the random motion driven by quantum field
fluctuations could become much more significant than the thermal
random motion. For example, in this limit, the velocity dispersion
of the charged test particle in the $z$-direction can be estimated
as
 \be
 \langle \Delta v_z^2\rangle\approx\;{e^2\over 4 \pi^2 m^2}{1\over
 z^2}+
{64e^2\pi^4\over 945m^2}\bigg({t\over \beta}\bigg)^2 \bigg({z\over
\beta}\bigg)^2{1\over \beta^2}+ {e^2\over 3\pi^2 m^2}{1\over t^2}
\;.
 \ee
 Clearly the first term represents the contribution of quantum
 vacuum fluctuations, while the second $\beta$ dependent term is
 the correction induced by system temperature being non-zero. With
 is result, it follows that in this case the ratio of the velocity dispersion
due to the random motion driven by quantum fluctuations to that
driven by thermal noise at the same temperature is
 \be
 {\Delta v_{qm}\over \Delta v_{th}
 }=\bigg({\alpha\over \pi}\bigg)^{1/2}\bigg({\beta\over
 z}\bigg)\;.
 \ee
This demonstrates that in the low temperature the random motion of
quantum origin can be orders of magnitude much more significant
than the thermal random motion and thus the quantum fluctuations
are the dominant driving source of the random of the test
particles at low temperature. To experimentally verify the
dominance of the random motion of the quantum origin over that of
thermal origin, one needs to cool the system to a significantly
low temperature, for example,  for $z\simeq 10^2 \mu m$, the
system temperature, $T$ has to be less than $0.1$ K. The smaller
the value of $z$, the lower the temperature $T$ has to be.

 In conclusion, we have been concerned with an interesting
problem of the random motion of charged test particles driven by
quantum electromagnetic field fluctuations at finite temperature.
Here, the random motion is caused by ever-existing quantum
electromagnetic fluctuations of a thermal bath of photons. A very
interesting feature of the random motion discussed in the present
paper, in contrast to that driven by thermal noise,  is that no
dissipation is needed for the velocity dispersion of the test
particle to be bounded at later times. Our calculations also show
that generally the random motion driven by quantum fluctuations is
one order of magnitude less significant than that driven by
thermal noise and it could be strengthened if the quantum field
fluctuations are to be modified by the presence of a reflecting
plane boundary. In particular, in the case with a reflecting plane
boundary,  the random motion of quantum origin in the direction
normal to the boundary could become orders of magnitude more
significant than that of thermal origin, when the system
temperature is low.

\begin{acknowledgments}
This work was supported in part  by the National Natural Science
Foundation of China  under Grants No.10375023 and No.10575035, the
Program for NCET (No.04-0784), the Key Project of Chinese Ministry
of Education (No.205110) and the Key Project of Hunan Provincial
Education Department (No. 04A030)

\end{acknowledgments}


\begin{thebibliography}{9}


\bibitem{Casimir}H. B. G. Casimir, Proc. K. Ned. Akad. Wet. {\bf51}, 793
(1948).

\bibitem{MR99} U. Mohideen and A. Roy. Phys. Rev. Lett., {\bf 83}, 3341 (1999).

\bibitem{SKL99} S. K. Lamoreaux. Phys. Rev. Lett., {\bf 84}, 5673 (2000) .

\bibitem{YF04} H. Yu and L.H. Ford, Phys. Rev. D {\bf 70}, 065009 (2004).

\bibitem{F95} L.H. Ford, Phys. Rev. D {\bf 51}, 1692 (1995).

\bibitem{YF99} H. Yu and L.H. Ford, Phys. Rev. D {\bf 60}, 084023 (1999).

\bibitem{YF00} H. Yu and L.H. Ford, Phys. Lett. B {\bf 496}, 107
(2000); gr-qc/0004063.

\bibitem{YW}H. Yu and P.X. Wu, Phys. Rev. D {\bf 68}, 084019 (2003).

\bibitem{YC} H. Yu and J. Chen, Phys. Rev. D {\bf 70}, 125006(2004).

\bibitem{EK} J.H. Eberly and A. Kujawski, Phys. Rev. {\bf 166},
197 (1968).

\bibitem{TY} M. Tan and H. Yu, Chin. Phys. Lett. {\bf 22}, 2165
(2005).

\bibitem{BM69} L.S. Brown and G.J. Maclay,
Phys. Rev. D {\bf 184}, 1272 (1969).




\end{thebibliography}
\end{document}